\documentclass[english,review,number,sort&compress]{elsarticle}
\usepackage[T1]{fontenc}
\usepackage[latin9]{inputenc}
\usepackage{geometry}
\usepackage{amsmath}
\usepackage{graphicx}
\usepackage{amssymb}

\newcommand{\lyxdot}{.}

\journal{Physics Letters A}

\begin{document}

\title{Outer Resonances and Effective Potential Analogy\\
in Two-Dimensional Dielectric Cavities}

\author[sgu]{Jinhang Cho}

\ead{locke@sogang.ac.kr}

\author[sgu]{Inbo Kim}

\author[sgu]{Sunghwan Rim}

\author[pcu]{Geo-Su Yim}

\author[sgu]{Chil-Min Kim\corref{cor1}}

\ead{chmkim@sogang.ac.kr}

\cortext[cor1]{Corresponding author. Tel.:+82 2 705 8428; Fax:+82 2 716 5374}

\address[sgu]{Acceleration Research Center for Quantum Chaos Applications, Sogang
University, Seoul 121-742, South Korea }

\address[pcu]{Department of Physics, Pai-Chai University, Daejeon 302-735, South
Korea }
\begin{abstract}
Outer resonances are studied as one type of quasinormal modes in two-dimensional
dielectric cavities with refractive index $n>1$. The outer resonances
can be verified as the resonances which survive only outside the cavity
in the small opening limit of the dielectric disk. We have confirmed
that the outer resonances universally exist in deformed cavities irrespective
of the geometry of cavity and they split into nearly degenerate states
in slightly deformed cavity. Also we have introduced an extended interpretation
of the effective potential analogy for the outer resonances. Since
most outer resonances in the dielectric cavities have quite high leakages,
they would affect to the broad background in the density of states.
But, for TE polarization case, relatively low-leaky outer resonances
exist and it presents the possibility that they can interact with
the inner resonances.\end{abstract}
\maketitle

\section{Introduction\label{sec:intro}}

Dielectric microcavities have been attracted much attention over the
past decade owing to their useful applicability as a prototype model
of mesoscopic open systems and a versatile element of hybrid optoelectronic
circuits \cite{Vahala}. The applicability of microcavities is based
on high quality factor achieved by so-called whispering gallery modes
(WGMs). The WGMs are formed in simple geometries with rotational symmetry
like spherical, cylindrical, and disk shapes resulting from complete
confinement of light by total internal reflection. But their isotropic
emission lacks directionality, so directional emission from slightly
deformed cavity has been intensively studied \cite{quadrupole_Noeckel,quadrupole_Gmachl,quadrupole_An,stadiumexp_Harayama,stadiumSBMexp_Harayama,triangle_Kurdoglyan,spiralexp_Chern,spiralexp_Courvoisier,limacon}.
Recently the issue of uni-directionality is resolved by the geometries
of spiral \cite{spiralexp_Chern,spiralexp_Courvoisier} and limaçon
\cite{limacon}.

Since the analytical analysis for the deformed microcavities combined
with quantum and classical aspects is almost impossible, various semiclassical
approaches have been applied to researches on their characteristics.
Scar theory \cite{scarbilliard_Heller}, which is originally developed
in the context of closed billiard system, is widely believed to be
suitable to explain the origin of the localized wave patterns of numerically
and experimentally observed resonances in microcavities \cite{quadrupole_Noeckel,quadrupole_Gmachl,quadrupole_An,stadiumexp_Harayama,scarcavity_SYLee,scarcavity_Wiersig}.
However, recent findings suggest that the scar-like localized resonance
patterns (i.e., quasiscar \cite{quasiscar_SYLee,quasiscar_TYKwon,Liu,quasiscar_CMKim})
are the effects of the openness of dielectric cavity rather than that
of unstable periodic orbits by the scar theory \cite{nearly_degen,Altmann}.
For this reason, it is necessary to examine closely the characteristics
of system according to the openness. Especially, our interest has
been focused on basic differences between open cavities and closed
billiards since a link between them remains a problem yet to be established.
As a part of the research on the openness of dielectric cavity, in
this report, we have investigated the outer resonances which do not
exist in the billiard systems and their characteristics through the
well-known effective potential analogy.

The paper is organized as follows: Firstly, we obtain the outer resonances
in two-dimensional dielectric circular disk and describe its properties
in Sec. \ref{sec:outer_disk}. In Sec. \ref{sec:Veff}, we apply the
effective potential analogy to the outer resonances and introduce
that extended interpretation is needed to describe the properties
of them. Finally, in Sec. \ref{sec:outer_spiral}, the general existence
of the outer resonances in two-dimensional cavities is verified through
the deformations of cavity shape.

\section{Outer resonances in Dielectric disk\label{sec:outer_disk}}

The solutions in the open system can be obtained in two perspectives
according to the boundary condition at infinity as follows: In scattering
perspective, the wavefunctions, called \textit{scattering states},
are composed of incoming plane waves and outgoing scattered waves.
The wavenumbers $k$ are real and are shown as the smoothly continuous
peak spectra structure. In emission perspective, the wavefunctions,
called \textit{resonances} or \textit{quasinormal modes} (QNMs), satisfy
the purely outgoing wave condition at infinity. The wavenumbers $k$
are complex with negative imaginary parts due to the leakage. The
real value and the imaginary value of $k$ correspond to wavelength
$\lambda$ ($\text{Re}[k]=2\pi/\lambda$) and lifetime $\tau$ ($\text{Im}[k]=-1/2c\tau$), respectively. The discrete wavenumbers on the complex space in
the emission perspective can be connected to the spectrum peaks in
the scattering perspective.

The Helmholtz equation of the dielectric disk as a typical model of
two-dimensional integrable open system is given by

\begin{equation}
\nabla^{2}\psi(k,r,\phi)+k^{2}\epsilon(r)\psi(k,r,\phi)=0,\label{eq:helmholtz-1}\end{equation}
\begin{equation}
\epsilon(r)=1+(n^{2}-1)\Theta(R-r),\label{eq:helmholtz-2}\end{equation}
where $k$ is the wavenumber, $\psi(k,r,\phi)$ is the wavefunction,
$n$ is the refractive index of the cavity, $R$ is the radius of
the disk, and $\Theta$ is the unit step function. We assume that
the refractive index outside the cavity is unity and $n>1$. On account
of the rotational symmetry, one can choose the solutions to be angular
momentum eigenstates. The exact wavefunctions for scattering states
are found to be

\begin{equation}
\psi_{m}(k,r,\phi)=\sqrt{\frac{k}{8\pi}}e^{-im\phi}\begin{cases}
I_{m}(k)\, J_{m}(nkr),\\
\quad\quad\quad\quad\quad\quad\quad0\leq r\leq R\\
H_{m}^{(2)}(kr)+S_{mm}(k)\, H_{m}^{(1)}(kr),\\
\quad\quad\quad\quad\quad\quad\quad r>R\end{cases}\label{eq:scattering_states}\end{equation}
where $m$ is angular momentum quantum number, $k$ is real wavenumber,
the $S$-matrix is diagonal in the angular momentum basis\begin{equation}
S_{lm}(k)=-\frac{H_{m}^{\prime(2)}(kR)-n\frac{J_{m}^{\prime}(nkR)}{J_{m}(nkR)}\, H_{m}^{(2)}(kR)}{H_{m}^{\prime(1)}(kR)-n\frac{J_{m}^{\prime}(nkR)}{J_{m}(nkR)}\, H_{m}^{(1)}(kR)}\,\delta_{lm},\label{eq:S-matrix}\end{equation}
and $I_{m}(k)$ is the mode strength amplitude \cite{Viviescas}.
The prime denotes the derivative with respect to $r$.

In Eq. \eqref{eq:scattering_states} for $r>R$, the Hankel function
of the second kind corresponds to an incident wave. To obtain the
QNMs in the emission perspective, we reduce Eq. \eqref{eq:scattering_states}
to be\begin{equation}
\psi_{m}(k,r,\phi)=\sqrt{\frac{k}{8\pi}}e^{-im\phi}\begin{cases}
I_{m}(k)\, J_{m}(nkr),\\
\quad\quad\quad\quad\quad\quad0\leq r\leq R\\
S_{mm}(k)\, H_{m}^{(1)}(kr),\\
\quad\quad\quad\quad\quad\quad r>R\end{cases}\label{eq:resonant_states}\end{equation}
because the QNMs satisfy purely outgoing boundary condition at infinity.
In this stage, we should note that the wavenumber $k$ is extended
from the real space to the complex space, i.e., the solution has a
leakage in the emission perspective.

Considering the boundary matching conditions for TM polarization at
$r=R$, we obtain the requirement\begin{equation}
\begin{array}{c}
I_{m}(k)\, J_{m}(nkR)=S_{mm}(k)\, H_{m}^{(1)}(kR)\\
I_{m}(k)\, nJ_{m}^{\prime}(nkR)=S_{mm}(k)\, H_{m}^{(1)\prime}(kR)\end{array}.\label{eq:BMC_requirement}\end{equation}
For having a non-trivial solution in the homogeneous system, the determinant\begin{equation}
D=\left|\begin{array}{cc}
J_{m}(nkR), & -H_{m}^{(1)}(kR)\\
nJ_{m}^{\prime}(nkR), & -H_{m}^{(1)\prime}(kR)\end{array}\right|\label{eq:determinant}\end{equation}
should be vanished. Using the recursion relations for Bessel and Hankel
functions of the first kind, the resonance condition is obtained as
follows,\begin{equation}
nJ_{m+1}(nkR)\, H_{m}^{(1)}(kR)=J_{m}(nkR)\, H_{m+1}^{(1)}(kR).\label{eq:BMC_TM}\end{equation}

By solving this equation, one can obtain complex wavenumbers $k_{r}$
for resonances and normalized wavefunctions\begin{equation}
\psi_{m}(k,r,\phi)=e^{-im\phi}\begin{cases}
A_{m}\, J_{m}(nk_{r}r), & 0\leq r\leq R\\
H_{m}^{(1)}(k_{r}r), & r>R\end{cases}\label{eq:normalized_wavefunction}\end{equation}
where $A_{m}$ is the normalized amplitude\begin{equation}
A_{m}\equiv\frac{I_{m}(k)}{S_{mm}(k)}=\frac{H_{m}^{(1)}(k_{r}R)}{J_{m}(nk_{r}R)}.\label{eq:normalized_amplitude}\end{equation}

We plot several resonances obtained numerically from the boundary
matching condition \eqref{eq:BMC_TM} for the dielectric disk with
a refractive index $n=2.0$ in Fig. \ref{fig:ResPos_Rek_vs_Imk_TM_TE}(a).
It is shown that the resonances are separated into two groups; One
group (black triangles) is composed of resonances with relatively
small absolute value of imaginary part and the other group (red circles)
have quite large absolute values of imaginary part.

Recently, Bogomolny \textit{et al}. obtained similar results in two-dimensional
dielectric disk with $n=1.5$. They separate these resonances into
two groups by the line of \begin{equation}
\text{Im}[kR]\sim-\frac{1}{2n}\ln\frac{n+1}{n-1}.\label{eq:imaginary_separation_value}\end{equation}
and called the low-leaky modes {}``internal resonances (Feshbach
resonances)'' and the high-leaky modes {}``external whispering gallery
modes'' or {}``outer resonances (shape resonances)'' \cite{Dubertrand,Bogomolny}.
We will call the low-leaky modes \textit{inner resonances} and the
high-leaky modes \textit{outer resonances} in this paper.

The wave intensity patterns of resonances and their sectional views
are illustrated in Fig. \ref{fig:Waves_TM_TE}. Figure \ref{fig:Waves_TM_TE}(a)
and \ref{fig:Waves_TM_TE}(b) are an inner resonance and an outer
resonance which have the same angular momentum quantum number $m=12$
and similar inside wavelength $\lambda_{in}\sim0.325$. Figure \ref{fig:Waves_TM_TE}(a)
corresponds to the eigenstate which has radial quantum number $l=2$
in the closed billiard system. But the mode in Fig. \ref{fig:Waves_TM_TE}(b)
can not be found in the corresponding billiard eigenstates and its
wave intensity inside the cavity is almost zero (not exactly zero).
Another outer resonance for $m=8$ is shown in Fig. \ref{fig:Waves_TM_TE}(c).
It also has similar wave intensity pattern to that for $m=12$, except
for the angular nodal numbers.

To investigate the correspondence to the closed system in more detail,
we checked up the tracing behavior of resonance mode as shown in Fig.
\ref{fig:ResTrace_TM}. In the limit $n\to\infty$ (small opening
limit), $k_{r}R$ of an inner resonance becomes zero but $nk_{r}R$
converges to some constant real value as shown in Figs. \ref{fig:ResTrace_TM}(a)
and \ref{fig:ResTrace_TM}(b). i.e., the mode becomes a bound state
without leakage and the wave is confined inside the cavity boundary.
Consequently, we can say that the bound state corresponds to a eigenmode
of the billiard with a specific real value $k$. 

But, strictly speaking, the billiard is not the same with the small
opening limit of cavity. Commonly used term, \textit{billiard} indicates
a non-leaky quantum system with Dirichlet boundary condition. For
inner resonance in TM polarization, a recent work shows that the real
part of $nk_{r}R$ for a given $m$ ($\neq0$) in the small opening
limit approaches zeros of $J_{m-1}(nk_{r}R)$ which are different
from the eigenvalues for the same $m$ in billiard \cite{Ryu}. Moreover,
it is recently reported that $nk_{r}R$ of an inner resonance with
$m=0$ corresponding to the ground state of billiard becomes zero
in the small opening limit \cite{Dettmann}. It has been known that
the modes disappearing in the closed system, so called {}``zero modes'',
exist. In Ref. \cite{OptProcess,zeromode_Leung}, the zero modes
were studied in one- and three-dimensional systems. We will deal with
the zero modes in two-dimensional cavity in another report.

In contrast to the tracing behavior of the inner resonance, $k_{r}R$
of the outer resonances is retained to some constant complex value
as $n\to\infty$. Dettmann \textit{et al}. show that $k_{r}R$ values
of outer resonances in the small opening limit become zeros of Hankel
function of the first kind \cite{Dettmann}. But $\text{Re}[nk_{r}R]$
of them diverges to $\infty$ (Figs. \ref{fig:ResTrace_TM}(c) and
\ref{fig:ResTrace_TM}(e)) and $\text{Im}[nk_{r}R]$ to $-\infty$
(Figs. \ref{fig:ResTrace_TM}(d) and \ref{fig:ResTrace_TM}(f)). i.e.,
the leakage of the modes inside the cavity becomes infinity. In Ref.
\cite{Dubertrand}, the {}``external whispering gallery modes''
arising from complex zeros of the Hankel functions were obtained in
the semiclassical limit by using Langer's formula. In their formulas,
we can also see that the imaginary part of $nk_{r}R$ becomes $-\infty$
in the limit $n\to\infty$. Thus the outer resonances in the small
opening limit exist only outside the cavity and can be considered
as the poles of scattering amplitude for the rigid cylinder \cite{Keller}.

For the case of TE polarization, we can obtain the resonance positions
by using the boundary matching condition \cite{thesis_Hentschel}
which is given as

\begin{eqnarray}
nJ_{m}(nkR)\, H_{m+1}^{(1)}(kR)-J_{m+1}(nkR)\, H_{m}^{(1)}(kR)\nonumber \\
=\frac{m}{kR}\left(n-\frac{1}{n}\right)J_{m}(nkR)\, H_{m}^{(1)}(kR).\label{eq:BMC_TE}\end{eqnarray}
The existence of outer resonances for TE polarization has been confirmed
by tracing the high-leaky modes represented by red circles in the
resonance position plot (Fig. \ref{fig:ResPos_Rek_vs_Imk_TM_TE}(b)).
Most of them are far from the inner resonances and the wave patterns
are similar to that of outer resonances in TM. But, differently from
TM case, we can see a group of the relatively low-leaky outer resonances
having $\text{Im}[k_{r}R]\sim-1.0$ and their inside wave intensities
are relatively strong as shown in Fig. \ref{fig:Waves_TM_TE}(h).
Some of them have been reported as {}``additional TE modes'' related
with the existence of the Brewster angle in Ref. \cite{Ryu}. The
existence of these low-leaky outer resonances has great significances
as follows: Firstly, density of states without the considering of
low-leaky outer resonances will bring a mismatch between theoretical
and experimental results. Secondly, inner resonances can interact
with the low-leaky outer resonances according to the change of system
parameters, since the low-leaky outer resonances are in the position
among the inner resonances as shown in Fig. \ref{fig:ResPos_Rek_vs_Imk_TE_n1.5}
when we just trace $k_{r}R$ of the resonances toward $n\to1.5$.
Therefore these outer resonances deserve to be considered in the study
of the resonance dynamics for inner resonances in TE case like avoided
resonance crossing or exceptional points.

\section{Effective potential analogy for Outer resonances\label{sec:Veff}}

The effective potential is a well-known analogy to explain the characteristics
of resonance modes in a symmetrical dielectric sphere or disk \cite{Johnson,thesis_Hentschel,thesis_Noeckel}.
The radial part of Helmholtz equation of the dielectric disk can be
written in the form of\begin{equation}
-\left[\frac{d^{2}}{dr^{2}}+\frac{1}{r}\frac{d}{dr}\right]\psi(r)+V_{eff}(r)\psi(r)=E\psi(r),\label{eq:radial_helmholtz}\end{equation}
where the effective potential is\begin{equation}
V_{eff}(r)=k^{2}\left[1-n^{2}(r)\right]+\frac{m^{2}}{r^{2}}.\label{eq:Veff}\end{equation}
 Here, the wavenumber $k$ and the energy $E$ defined as $k^{2}$
are real because the analogy is rooted to the scattering perspective.
Due to the interplay between the dielectric potential with refractive
index and the repulsive centrifugal potential, the effective potential
has the form of metastable well as shown in Fig. \ref{fig:Veff_Inner_Outers_TM}(a).
The classical turning points are defined by the condition $E-V_{eff}(r)=k^{2}n^{2}-m^{2}/r^{2}=0$
and a classically allowed or classically forbidden region is represented
as positive or negative value of $E-V_{eff}(r)$, respectively. Using
this condition, turning points on the boundary lead the relations
for the maximum and the minimum possible value of $k$ that the wave
can be trapped in the well for a given $n$ and $m$ as follows,\begin{equation}
k_{T}^{2}=\left(\frac{m}{R}\right)^{2},\label{eq:maximum_possible_value}\end{equation}
\begin{equation}
k_{B}^{2}=\left(\frac{m}{nR}\right)^{2},\label{eq:minimum_possible_value}\end{equation}
where $R$ is the radius of the disk.

For a given $m$, the top of the potential well at the disk boundary
$R=1.0$ is fixed to $k_{T}^{2}$ and independent of the variation
of $k$ and $n$. The bottom of the potential well at the boundary
meets with $k_{B}^{2}$ when $E=k_{B}^{2}$ and it has the dependence
on $k$ and $n$. If $k$ is fixed and $n$ increases, the depth of
the potential well may further deepen. In the case that $n$ is fixed,
if $k^{2}$ is larger than $k_{B}^{2}$, the bottom moves downward.
On the contrary, if $k^{2}$ becomes smaller than $k_{B}^{2}$, it
moves upward, and eventually the potential well becomes very shallow.

Figure \ref{fig:ResPos_m_vs_Rek_TM_TE}(a) shows the inner and the
outer resonance positions for $\text{Re}[k_{r}R]$ vs $m$. One can
easily find the inner resonances above the dotted line $\text{Re}[kR]=m$,
namely \textit{above-barrier resonances} and the inner resonances
in the trap range ($m/n<\text{Re}[kR]<m$), namely \textit{below-barrier
resonances} \cite{thesis_Hentschel,thesis_Noeckel}. They are represented
by black triangles in Fig. \ref{fig:ResPos_m_vs_Rek_TM_TE}. In general,
wave oscillates in classically allowed region and diffuses in classically
forbidden region. For the below-barrier resonances, the waves are
well trapped in the potential well. The below-barrier resonances decay
only by tunneling via the effective potential barrier and the absolute
values of $\text{Im}[k_{r}R]$ are very small, i.e., high-Q modes
are formed. But we must take notice that the absolute values of imaginary
part of the above-barrier resonances are also comparatively small
although they are in non-trap region, because the potential depth
is more deepen as $\text{Re}[k_{r}R]$ increase.

In contrast to the inner resonances, the outer resonances have extremely
large absolute values of $\text{Im}[k_{r}R]$. We have verified the
probability densities of the imaginary parts of the outer resonances,
the above-barrier resonances, and the below-barrier resonances obtained
in the region of $m\leq20$ and $\text{Re}[k_{r}R]\leq20.0$. The
aspects of distributions for three groups are certainly different.
The below-barrier resonances are concentrated in the region of $\left|\text{Im}[kR]\right|\sim0.025$
and the above-barrier resonances are concentrated in the region of
$\left|\text{Im}[kR]\right|$ about 10 times as large as that of below-barrier
resonances. The absolute values of imaginary parts of the outer resonances
are over 10 times as large as that of above-barrier resonances. In
the closed system, the density of states is represented in the forms
of discrete series of delta peaks positioned at the eigenvalues $k$.
While, in the case of an open cavity, the peaks positioned at the
real parts of resonance spectra $k_{r}R$ acquire some widths which
are related with the imaginary parts of $k_{r}R$. As $\left|\text{Im}[k_{r}R]\right|$
increases, the peak becomes broader. Generally, the density of states
in a cavity is composed of not only sharp peaks corresponding to below-barrier
resonances but also broad band corresponding to above-barrier resonances.
The effects of very broad peaks of the outer resonances, which have
large absolute values of imaginary part, are also contained in the
background.

The outer resonances in Fig. \ref{fig:ResPos_Rek_vs_Imk_TM_TE}(a)
are represented by red circles in Fig. \ref{fig:ResPos_m_vs_Rek_TM_TE}(a).
For a given $m$, the number of the outer resonances can be estimated
at $[m/2]$. Most of the outer resonances exist under the line of
$k_{B}$ and some of them for the case of $m\geq4$ are partially
located in the trap region. We have drawn the effective potentials
for two of outer resonances with $\text{Re}[k_{r}R]$ for $m=12$
($\text{Re}[k_{r}R]=9.67774$) and $m=8$ ($\text{Re}[k_{r}R]=2.20483$)
in Fig. \ref{fig:Veff_Inner_Outers_TM}(b) and \ref{fig:Veff_Inner_Outers_TM}(c).
In the case of the mode for $m=8$, the wave inside the boundary exists
in classically forbidden region as shown in Fig. \ref{fig:Veff_Inner_Outers_TM}(c)
and the intensity of the wavefunction inside the cavity is nearly
zero as shown in Fig. \ref{fig:Waves_TM_TE}(g). Hence we are apt
to naively think that the effective potential analogy agrees with
the wave pattern. However, for the case of $m=12$ as shown in Fig.
\ref{fig:Veff_Inner_Outers_TM}(b), it seems as if the wave inside
the cavity should be trapped in the well. According to the effective
potential analogy, the wave located in the trap region like Fig. \ref{fig:Veff_Inner_Outers_TM}(a)
lives for a long time inside the cavity and its inside intensity is
strongly localized in the classically allowed region as shown in Fig.
\ref{fig:Waves_TM_TE}(e). But, in fact as shown in Fig. \ref{fig:Waves_TM_TE}(f),
the outer resonance for $m=12$ has very low inside intensity similar
to that of $m=8$, even though the resonance is in the trap region
and has the wavelength $\lambda_{in}=0.32462$ smaller than the length
of the effective potential well $l_{w}=0.38002$. So we can conclude
that the analogy can not be applied to the outer resonances and we
need a different description of the effective potential. If we suppose
that inner and outer resonances are formed inside and outside of the
cavity, respectively, it can be interpreted as that the intensity
of an outer resonance formed with $E$ between the cavity boundary
and infinite boundary rapidly decrease inward the cavity by tunneling
via the classically forbidden region of potential barrier. Therefore
it does not matter whether the energy of the outer resonance is in
the trap region or not.

Such description of effective potential for outer resonances can be
more clarified in the case of TE polarization. The effective potential
analogy in two-dimensional dielectric disk can be similarly applied
to TE case with one difference that $\psi(r)$ is the magnetic field
instead of the electric field. In Fig. \ref{fig:ResPos_m_vs_Rek_TM_TE}(b),
we obtain the resonance positions for $\text{Re}[k_{r}R]$ vs $m$
in the case of TE polarization. Differently from TM case, the number
of the outer resonances for a given $m$ can be estimated at $[(m+1)/2]$
and we can see the outer resonances located above $k_{T}$-line for
$m\geq3$ and below $k_{T}$-line for $m\leq2$, which are the relatively
low-leaky outer resonances in Fig. \ref{fig:ResPos_Rek_vs_Imk_TM_TE}(b).
$E$ of these low-leaky outer resonances existing near the $k_{T}$-line
is above the potential barrier or has thin potential barrier than
that of other outer resonances. So they can more easily tunnel in
the cavity and, as a result, have relatively large inside intensity.
Figure \ref{fig:VeffW_LlOuter_TE} shows the variation of wavefunction
of a low-leaky outer resonance for $m=17$ when effective potential
is changed by $n$. $\sqrt{E}$ of the resonance is changed from (a)
$18.94011$ (over the line of $k_{T}$) to (b) $16.38559$ (under
the line of $k_{T}$). This figure well represents that the relatively
strong inside intensity with oscillatory property at $n=2.0$ becomes
extremely small as the energy located below the tip of the potential
barrier ($k_{T}^{2}$) at $n=9.0$.

\section{Outer Resonances in Deformed cavities\label{sec:outer_spiral}}

In order to show that the existence of outer resonances are generic
feature of the two-dimensional cavities, we investigated the outer
resonances in a spiral-shaped cavity. It is a classically complete
chaotic system and has unique features different from typical deformed
cavities (e.g., totally asymmetrical geometry and the presence of
a notch) \cite{spiralexp_Chern,spiralexp_Courvoisier,quasiscar_SYLee,quasiscar_TYKwon}.
The spiral cavity boundary is given by\begin{equation}
r(\phi)=R\left(1+\epsilon\frac{\phi}{2\pi}\right)\label{eq:spiral_geometry}\end{equation}
in polar coordinates $(r,\phi)$, with the radius of spiral $R=1.0$
at $\phi=0$, the deformation parameter $\epsilon$, and the refractive
index $n$.

The resonance positions and wave patterns in deformed cavities can
be obtained with the boundary element method (BEM) \cite{BEM}. We
have found the the spots of high-leaky solutions distributed on the
region of $\text{Im}[kR]<-1.5$, where $\epsilon=0.1$ and $n=3.0$.
They have been verified as the outer resonances and two nearly degenerate
states of them are illustrated in Fig. \ref{fig:Outer_Waves_deformed_TM}(a)
and \ref{fig:Outer_Waves_deformed_TM}(b). The outer resonances in
the dielectric disk split into nearly degenerate states in the spiral
cavity by shape perturbation. The splitting of a degenerate mode for
$m\neq0$ is a general aspect in the cavities slightly deformed from
a perfect symmetric geometry like the dielectric disk. Especially,
in the spiral-shaped cavity, the pairs of nearly degenerate modes
have different wavelengths and $Q$-factors due to the broken chirality
\cite{spiral_degen_pair}.

For more confirmation, we obtained the resonance positions and the
resonance patterns in low-$nkR$ regime for a square-shaped cavity
and a stadium-shaped cavity and easily found the outer resonances
in there too as shown in Fig. \ref{fig:Outer_Waves_deformed_TM}(c)
and \ref{fig:Outer_Waves_deformed_TM}(d). Hereby, we can see that
the outer resonances universally exist as one group of QNMs in two-dimensional
dielectric cavities whether the geometry of cavity is integrable or
chaotic.

\section{Conclusions\label{sec:conclusions}}

In the dielectric disk with $n>1$ as a two-dimensional open system,
we established the existence of outer resonances besides the inner
resonances. The outer resonances do not have the counterparts in billiard
and disappear inside the cavity due to the infinite leakage in the
small opening limit. For the finite refractive index, most of them
are very short-lived resonances and the wave intensities are nearly
zero inside the cavity. But, exceptionally, the outer resonances for
$\text{Re}[k_{r}R]\gtrsim m$ in the TE case are relatively long-lived
resonances and exist around the inner resonances in the case of low
refractive index. It presents the possibility that the outer resonances
can interact with the inner resonances.

Using BEM, we showed that the outer resonances universally exist as
one group of QNMs in two-dimensional dielectric cavities irrespective
of the geometry of cavity and, especially, exist as the nearly degenerate
states in the slightly deformed cavity. The outer resonances constitute
extremely broad background in the density of states because of their
quite high leakage and they should be take into account in the study
of the trace formula in open cavity \cite{Bogomolny,Jain}.

Also, it is known that the effective potential analogy arising from
the scattering perspective can be well applied for the description
of the QNMs in the dielectric disk. However, we noticed that care
must be taken to interpret effective potential depending on the class
of resonances.

\section*{Acknowledgment\label{sec:ack}}

We would like to thank J.-W. Ryu and S.-Y. Lee for discussions. This
work was supported by Acceleration Research (Center for Quantum Chaos
Applications) of MEST/KOSEF.

\newpage{}

\begin{center}
\textbf{\large Figure Captions}
\par\end{center}{\large \par}

\textbf{Figure \ref{fig:ResPos_Rek_vs_Imk_TM_TE}:} Resonance positions
in the complex $kR$ space obtained from the boundary matching conditions
for (a) TM and (b) TE polarizations in the dielectric disk with $n=2.0$.
The resonances are separated into two groups by a dashed line. One
group (triangles) is composed of resonances with relatively small
absolute value of imaginary part (inner resonances), the other group
(circles) has quite large absolute value of imaginary part (outer
resonances).

\begin{figure}
\begin{centering}
\includegraphics[width=4cm]{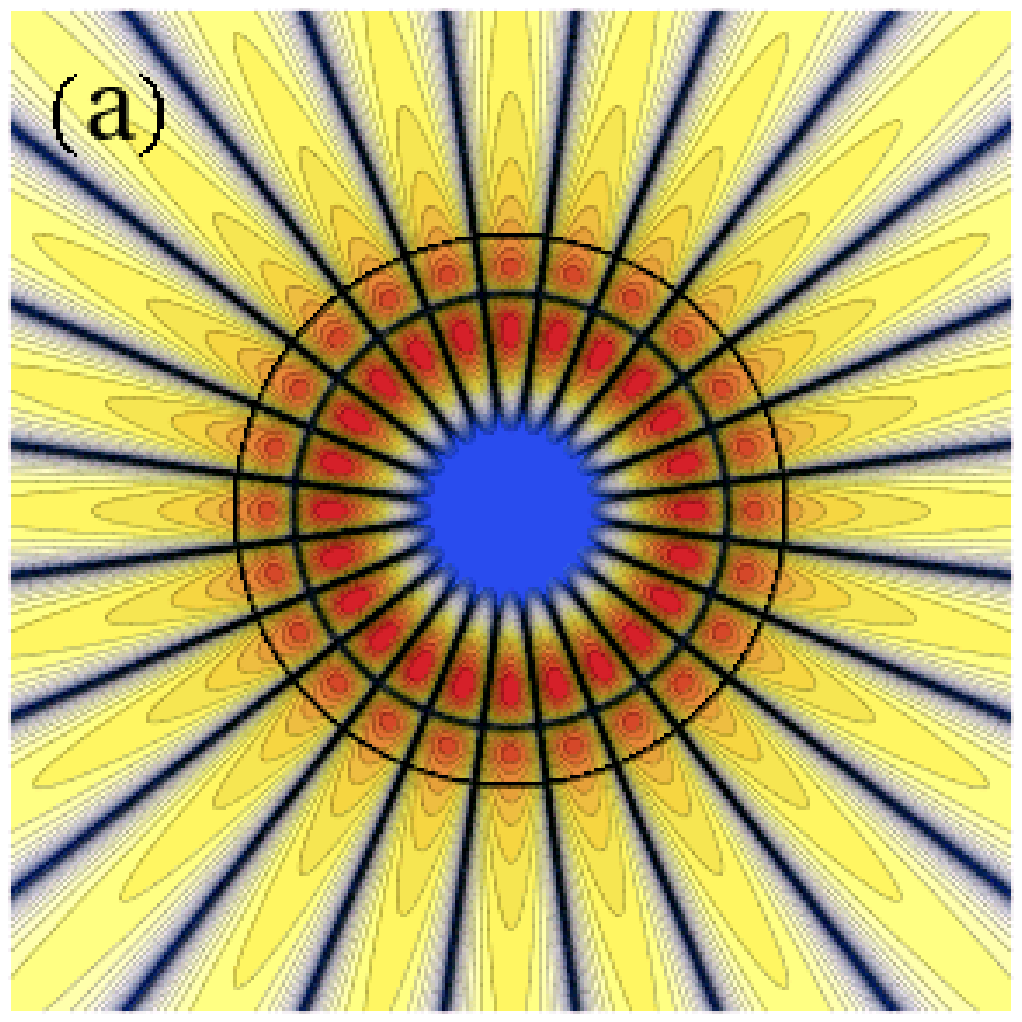}\includegraphics[width=4cm]{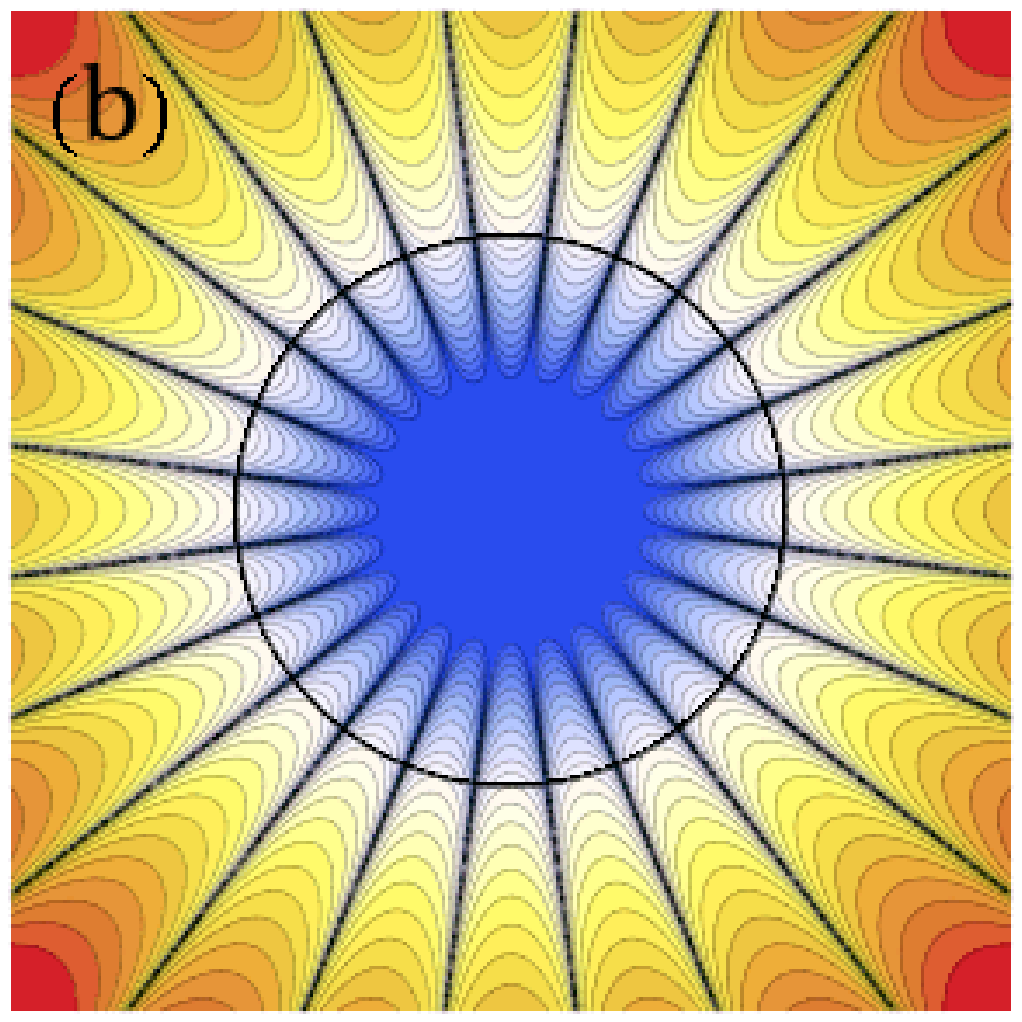}\includegraphics[width=4cm]{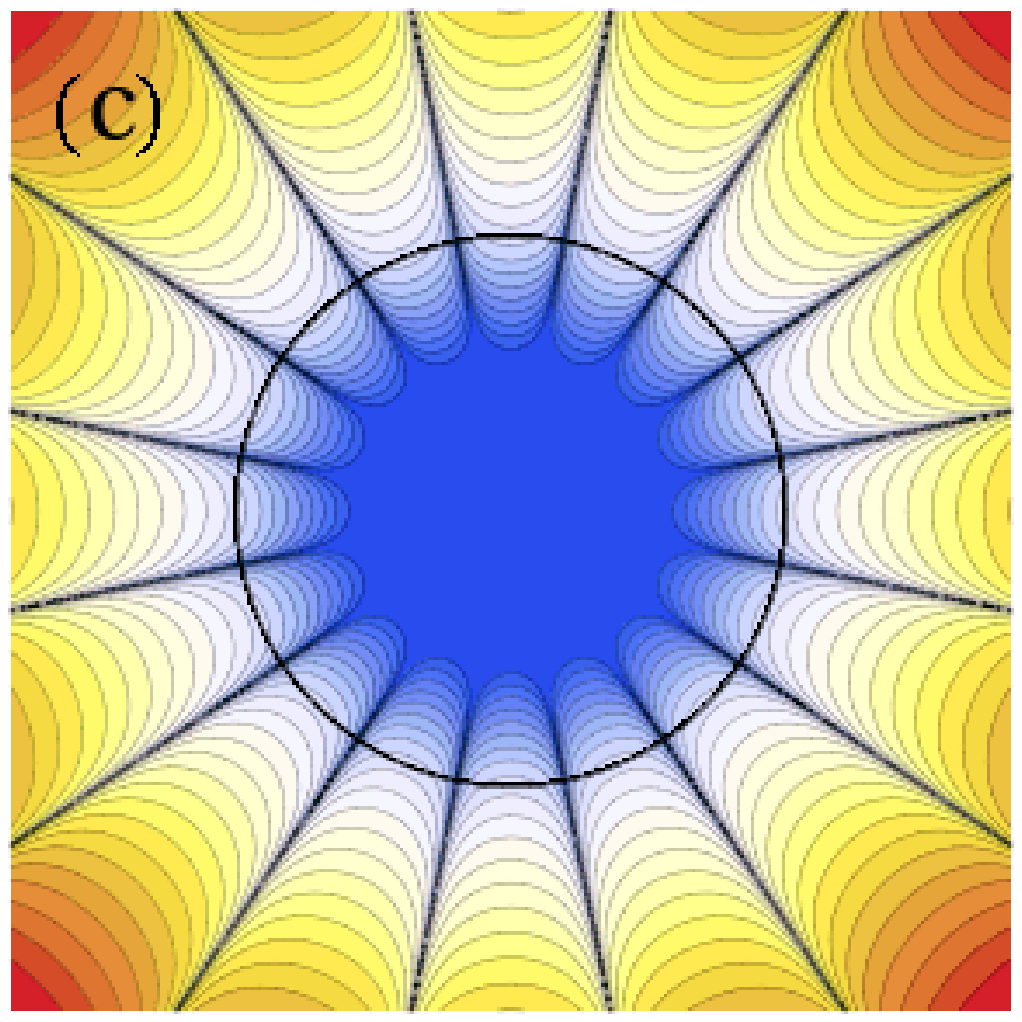}\includegraphics[width=4cm]{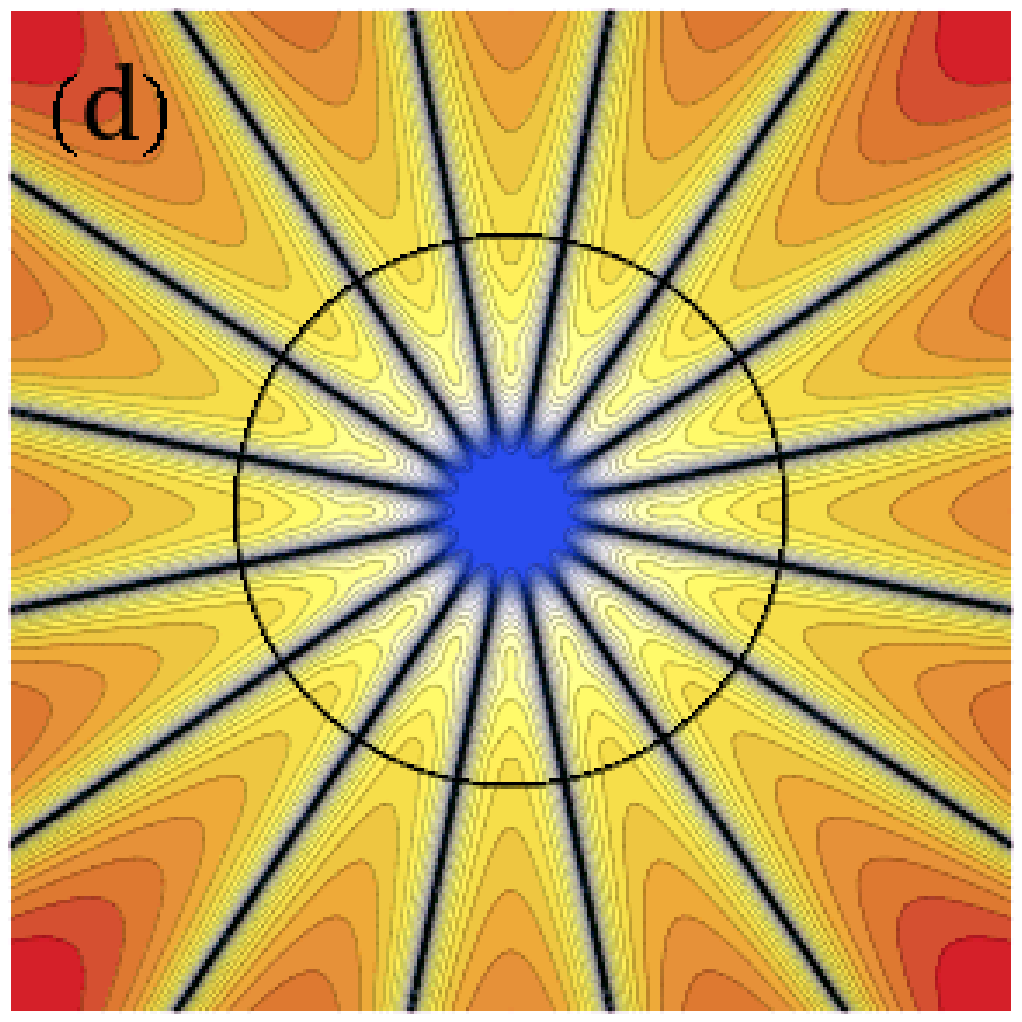}
\par\end{centering}

\centering{}\includegraphics[width=4cm]{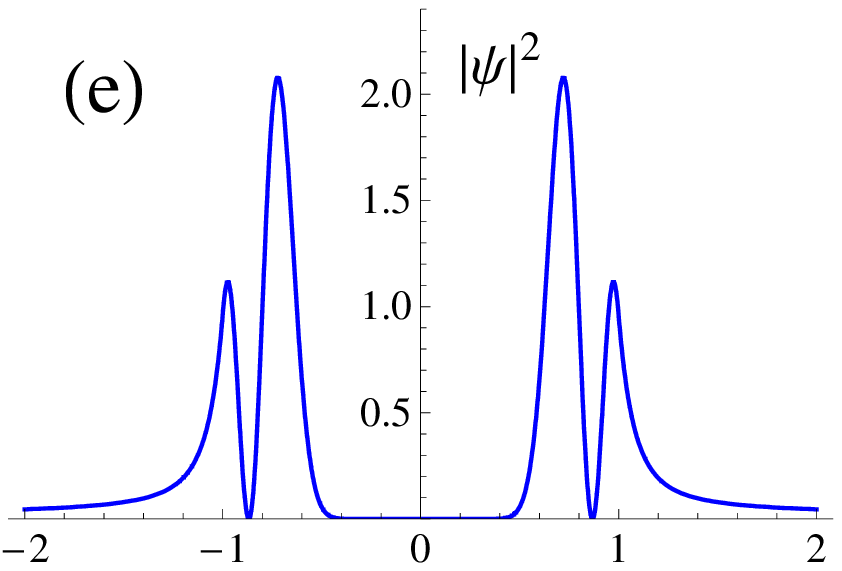}\includegraphics[width=4cm]{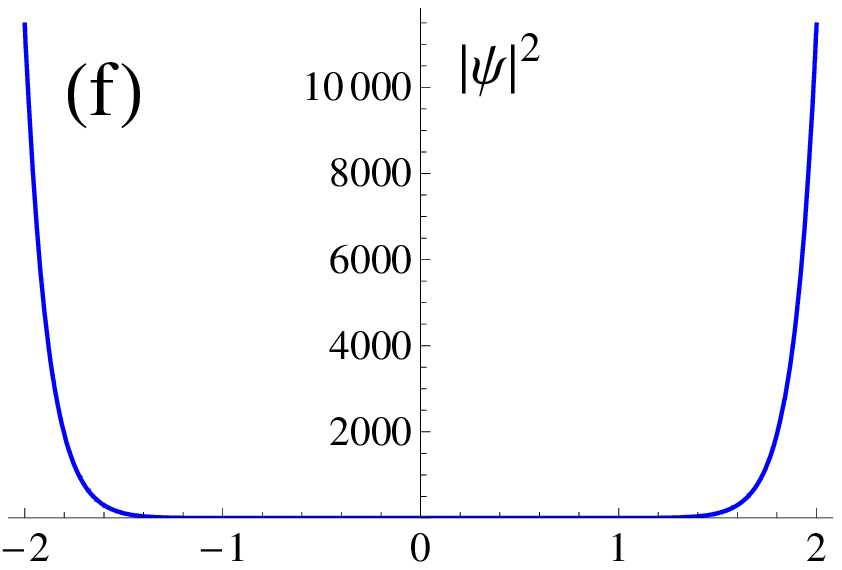}\includegraphics[width=4cm]{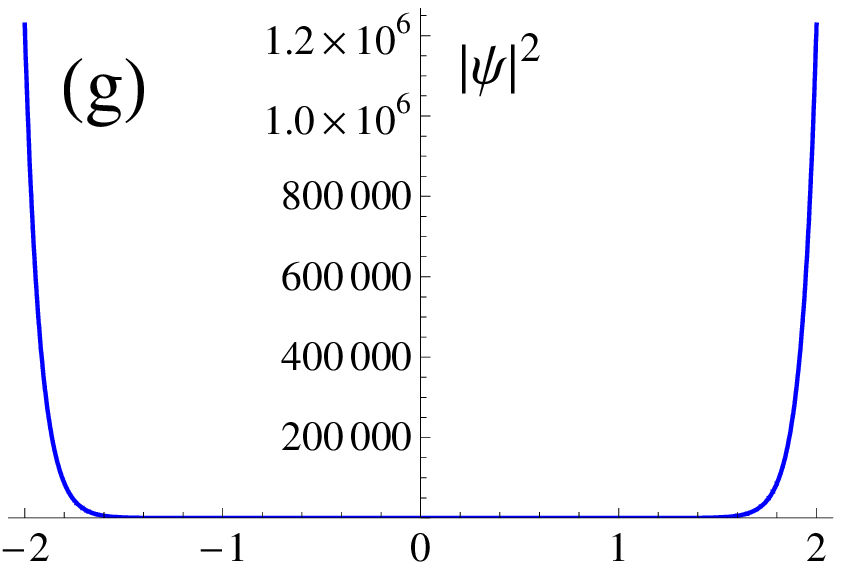}\includegraphics[width=4cm]{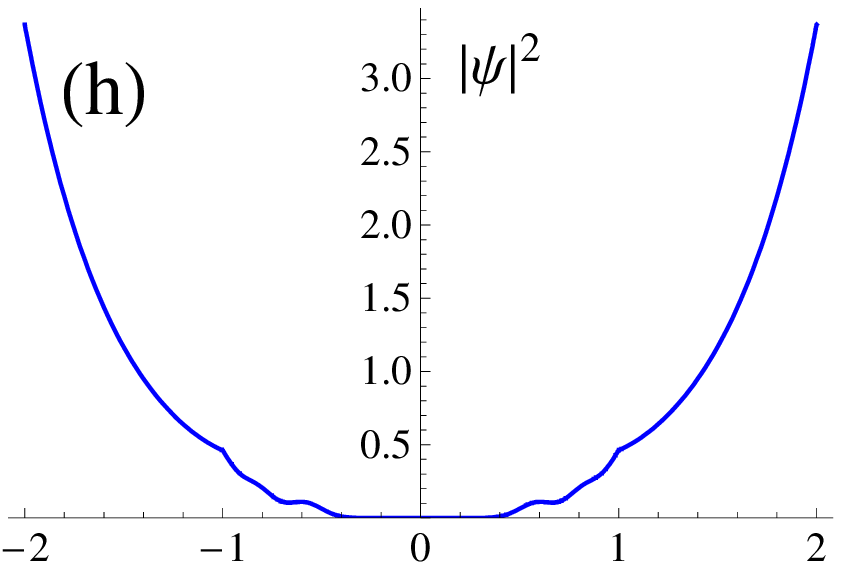}\caption{\label{fig:Waves_TM_TE}}

\end{figure}

\textbf{Figure \ref{fig:Waves_TM_TE}:} Wave patterns of (a) an inner
resonance for $m=12$, $k_{r}R=9.61998-i0.02300$ (TM) and three outer
resonance for (b) $m=12$, $k_{r}R=9.67774-i3.86429$ (TM) and (c)
$m=8$, $k_{r}R=2.20483-i5.50279$ (TM), and (d) $m=8$, $k_{r}R=8.67533-i1.24009$
(TE) in the dielectric disk with $n=2.0$. (e), (f), (g), and (h)
are the sectional views of (a), (b), (c), and (d), respectively.

\textbf{Figure \ref{fig:ResTrace_TM}:} Mode tracing behavior for
an inner resonance and two outer resonance in the dielectric disk
for TM polarization. Tracing $\text{Re}[nk_{r}R]$ and $\text{Im}[nk_{r}R]$
of (a), (b) inner resonance for $m=12$, $l=2$, (c), (d) outer resonance
for $m=12$, and (e), (f) outer resonance for $m=8$, respectively.

\textbf{Figure \ref{fig:ResPos_Rek_vs_Imk_TE_n1.5}:} Resonance positions
in the complex $kR$ space obtained from the boundary matching condition
for TE polarization in the dielectric disk with $n=1.5$. The triangular
points are inner resonances and the circular points are outer resonances.

\begin{figure}
\centering{}\includegraphics[width=2.8cm]{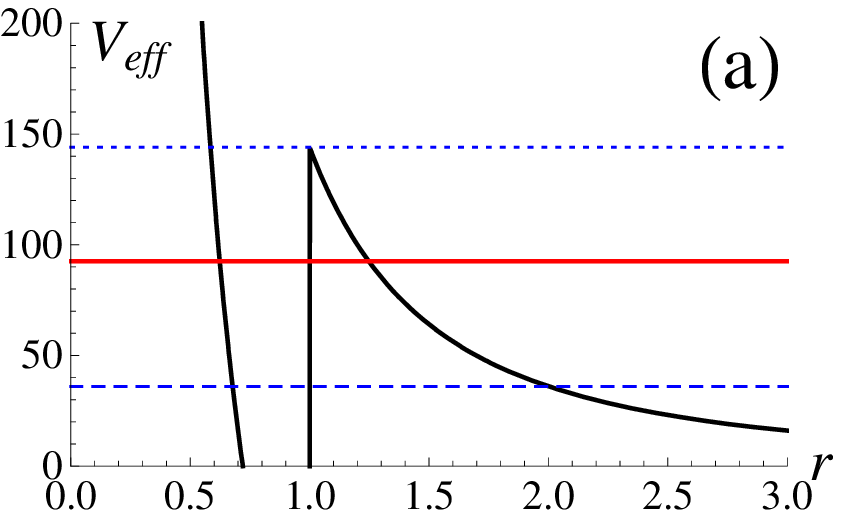}\includegraphics[width=2.8cm]{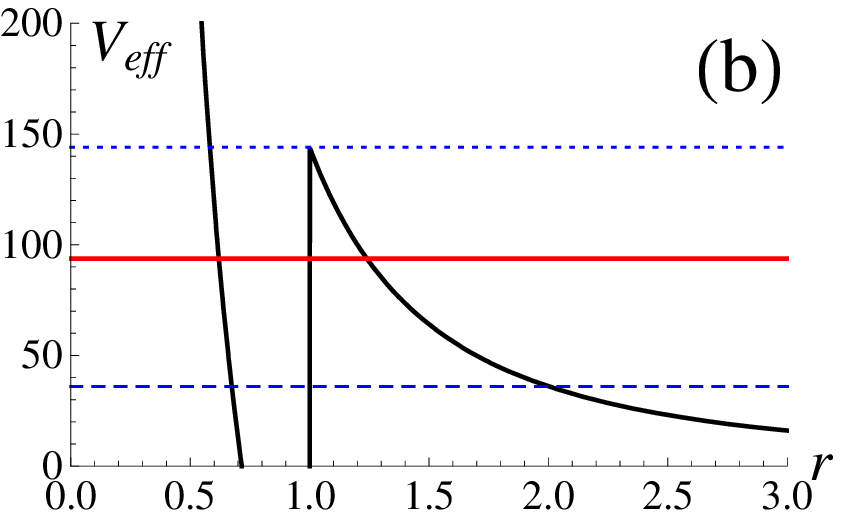}\includegraphics[width=2.8cm]{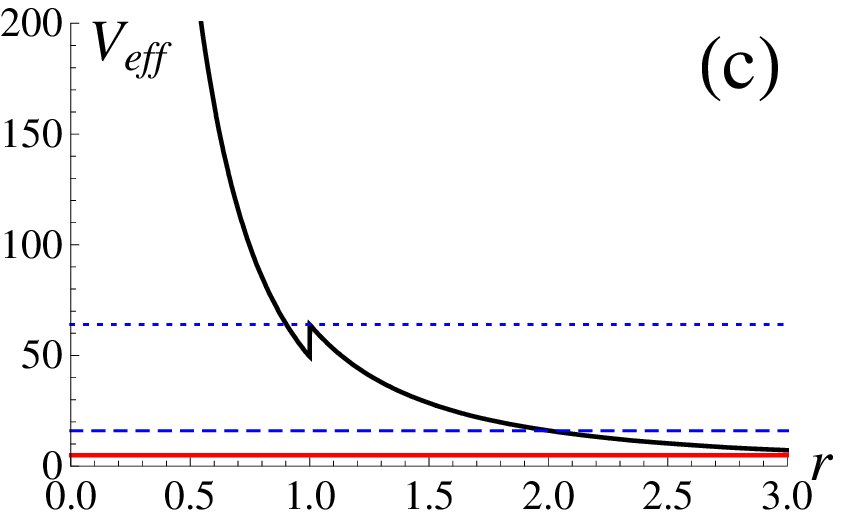}\caption{\label{fig:Veff_Inner_Outers_TM}}

\end{figure}

\textbf{Figure \ref{fig:Veff_Inner_Outers_TM}:} Effective potential
for a general inner resonance for (a) $m=12$, $l=2$ ($\text{Re}[k_{r}R]=9.61998$)
and outer resonances for (b) $m=12$, $\text{Re}[k_{r}R]=9.67774$
and (c) $m=8$, $\text{Re}[k_{r}R]=2.20483$ in the dielectric disk
for TM polarization, where $n=2.0$ and $R=1.0$. Red solid line is
$\text{Re}[k_{r}]^{2}=E$, blue dotted line is $k_{T}^{2}$, and blue
dashed line is $k_{B}^{2}$.

\textbf{Figure \ref{fig:ResPos_m_vs_Rek_TM_TE}:} Resonance positions
for $\text{Re}[k_{r}R]$ vs $m$ obtained from the boundary matching
conditions for (a) TM and (b) TE polarizations in the dielectric disk
with $n=2.0$. Inner resonances and outer resonances are represented
by triangular points and circular points, respectively. Dotted line
is $k_{T}$ of effective potential well ($\text{Re}[kR]=m$) and dashed
line is $k_{B}$ ($\text{Re}[kR]=m/n$).

\begin{figure}
\begin{centering}
\includegraphics[width=4.3cm]{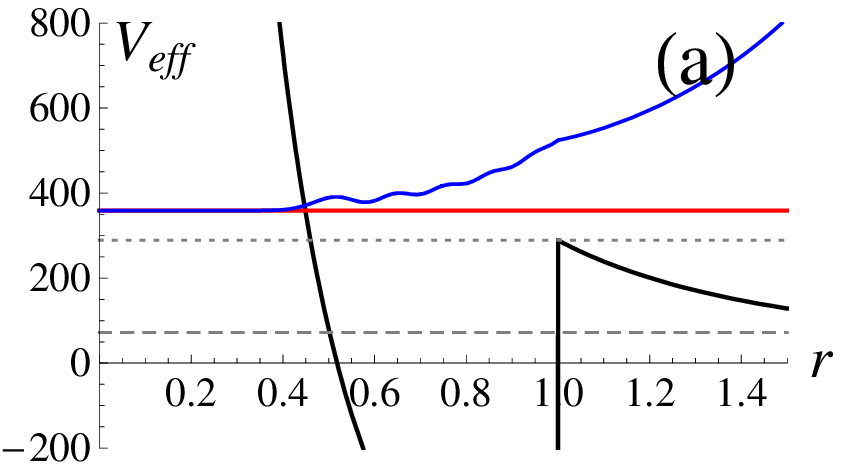}\includegraphics[width=4.3cm]{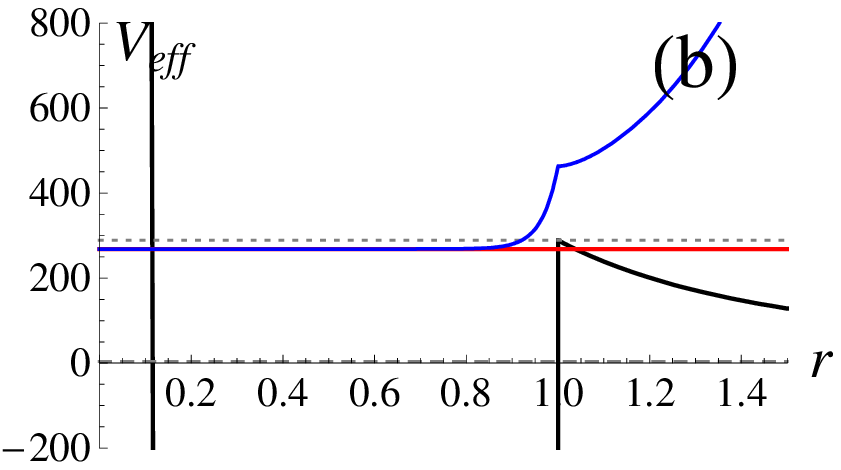}
\par\end{centering}

\caption{\label{fig:VeffW_LlOuter_TE}}

\end{figure}

\textbf{Figure \ref{fig:VeffW_LlOuter_TE}:} Effective potentials
and wavefunctions for a low-leaky outer resonance ($m=17$) in the
dielectric disk with (a) $n=2.0$ ($k_{r}R=18.94011-i1.30632$) and
(b) $n=9.0$ ($k_{r}R=16.38559-i1.62816$). Black, red, and blue solid
lines are $V_{eff}$, $E$, and $\left|\psi(r)\right|^{2}$ (arbitrary
unit), respectively.

\begin{figure}
\begin{centering}
\includegraphics[width=8.5cm]{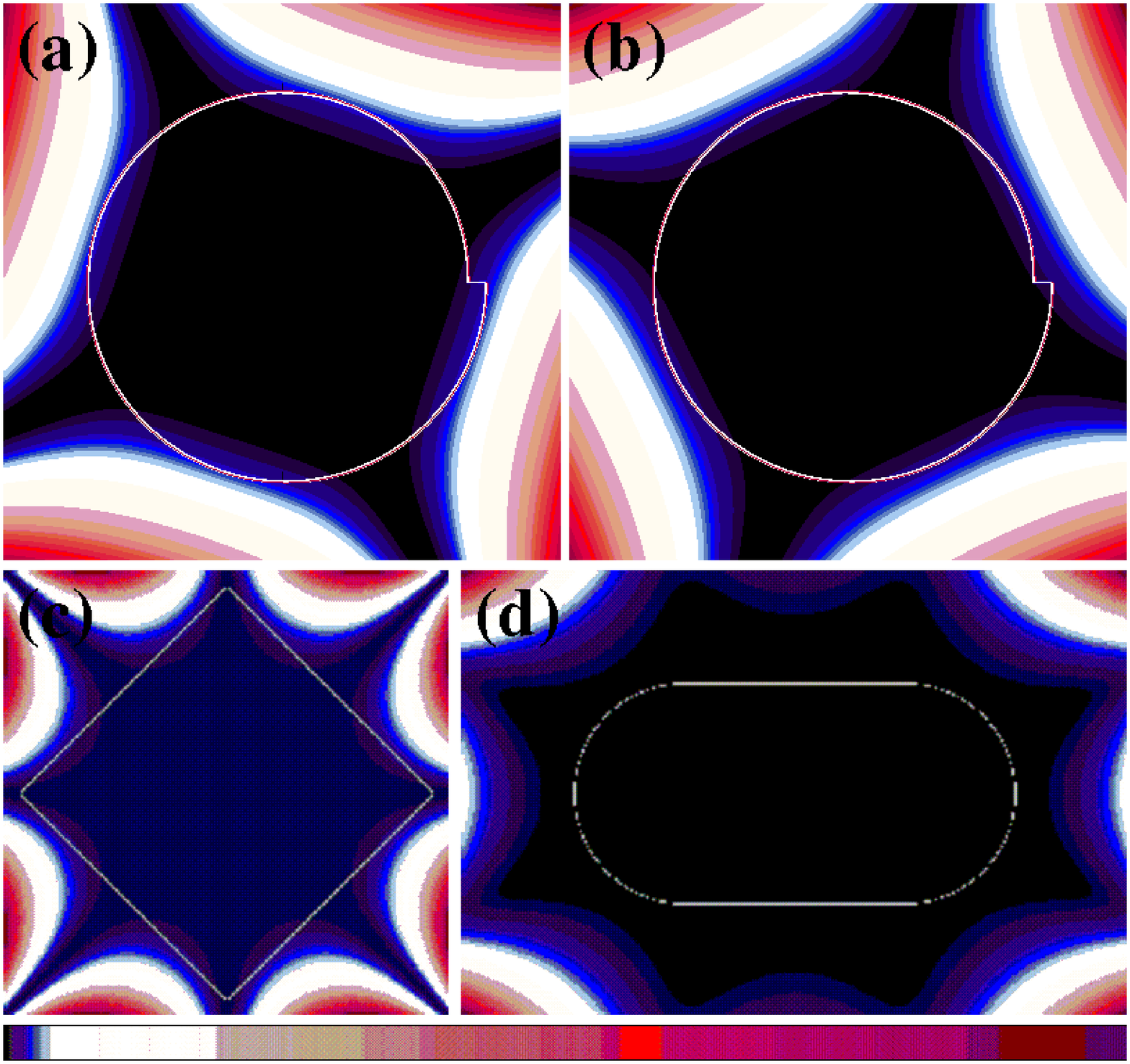}
\par\end{centering}

\centering{}\caption{\label{fig:Outer_Waves_deformed_TM}}

\end{figure}

\textbf{Figure \ref{fig:Outer_Waves_deformed_TM}:} Wave patterns
for outer resonances in deformed cavities for TM polarization. (a)
$k_{r}R=0.41394-i1.53316$ and (b) $k_{r}R=0.41708-i1.54480$ are
nearly degenerated modes in the spiral-shaped cavity with $\epsilon=0.1$
where $n=3.0$, respectively. (c) $k_{r}R=1.97536-i2.25493$ and (d)
$k_{r}R=1.43900-i1.55748$ are the outer resonances in the square-shaped
cavity and the stadium-shaped cavity with deformation parameter $L/R=1.0$
where $n=2.0$, respectively.

\newpage{}%
\begin{figure}
\centering{}\includegraphics[width=4cm]{./ResPos_k_TM}\includegraphics[width=4cm]{./ResPos_k_TE}\caption{\label{fig:ResPos_Rek_vs_Imk_TM_TE}}

\end{figure}
\begin{figure}
\centering{}\includegraphics[width=8cm]{./trace}\caption{\label{fig:ResTrace_TM}}

\end{figure}
\begin{figure}
\begin{centering}
\includegraphics[width=8cm]{./ResPos_k_TE_n1\lyxdot 5}
\par\end{centering}

\caption{\label{fig:ResPos_Rek_vs_Imk_TE_n1.5}}

\end{figure}
\begin{figure}
\centering{}\includegraphics[width=4cm]{./ResPos_m_TM}\includegraphics[width=4cm]{./ResPos_m_TE}\caption{\label{fig:ResPos_m_vs_Rek_TM_TE}}

\end{figure}

\end{document}